**Realization of high-dynamic-range broadband magnetic-field sensing with ensemble nitrogen-vacancy centers in diamond**

Cao Wang,[1,2] Qihui Liu,[1,2] Yuqiang Hu,[3,4] Fei Xie,[1,2] Krishangi Krishna,[5] Nan Wang,[1,2] Lihao Wang,[1] Yang Wang,[1] Kimani C. Toussaint Jr,[5] Jiangong Cheng,[1,a)] Hao Chen,[1,a)] and Zhenyu Wu[1,2,3,4,a)]

[1]*State Key Laboratory of Transducer Technology, Shanghai Institute of Microsystem and Information Technology, Chinese Academy of Sciences, Shanghai, 200050, China*

[2]*University of Chinese Academy of Sciences, Beijing, 100049, China*

[3]*School of Microelectronics, Shanghai University, 200444 Shanghai, China*

[4]*Shanghai Industrial uTechnology Research Institute, 201800 Shanghai, China*

[5]*School of Engineering, Brown University, Providence, RI 02912, United States*

[a]*Authors to whom correspondence should be addressed: jgcheng@mail.sim.ac.cn, haochen@mail.sim.ac.cn and zhenyu.wu@ mail.sim.ac.cn*

We present a new magnetometry method integrating an ensemble of nitrogen-vacancy (NV) centers in a single-crystal diamond with an extended dynamic range for monitoring the fast changing magnetic-field. The NV-center spin resonance frequency is tracked using a closed-loop frequency locked technique with fast frequency hopping to achieve a 10 kHz measurement bandwidth, thus, allowing for the detection of fast changing magnetic signals up to 0.723 T/s. This technique exhibits an extended dynamic range subjected to the working bandwidth of the microwave source. This extended dynamic range can reach up to 4.3 mT, which is 86 times broader than the intrinsic dynamic range. The essential components for NV spin control and signal processing such as signal generation, microwave frequency control, data processing and readout are integrated in a board-level system. With this platform, we demonstrate broadband magnetometry with an optimized sensitivity of 4.2 nT·Hz$^{-1/2}$. This magnetometry method has the potential to be implemented in a multichannel frequency locked vector magnetometer suitable for a wide range of practical applications such as magnetocardiography and high-precision current sensors.

## I. INTRODUCTION

Magnetometers with nitrogen-vacancy (NV) color centers in diamonds, alternatively known as solid-state quantum sensors, continue to attract attention primarily due to their extraordinarily high sensitivity, high spatial resolution, temperature self-calibration, wide-temperature-range operating conditions, and inherent vector capabilities.

[1-4] A combination of these characteristics can be exploited for a myriad of applications, including magnetocardiography,[5] neuron activity sensing,[6,7] nuclear magnetic resonance spectroscopy,[8-10] earth magnetic field anomalies,[11] magnetic current imaging of an integrated circuit,[12] strain imaging,[13] electrometry,[14] and velocimetry.[15] Optically detected magnetic resonance (ODMR) combined with frequency modulation and demodulation techniques is the most robust and general spin readout method. This is often used as a magnetic sensing protocol for quantum sensing based on a single or an ensemble of NV centers. [6,16-18]

At present, most NV-based magnetometry systems for the control and detection of electron spin states still rely on bulky table-top equipment,[6,17,19,20] and recent advances in miniaturizing lab-based magnetometers using partially integrated demodulation systems have been reported.[18,20-22] However, the miniaturization of conventional quantum sensing platforms still remains a challenge. Moreover, innovations involving the real-time measurement of external magnetic fields at high tracking rates with NV-diamond sensors are rare, particularly when the dynamic range goes beyond the intrinsic range of the NV resonance. A high tracking rate and dynamic range is required to monitor the abrupt field changes covering a broad range for practical applications, especially in current sensing realm.[23] Although feedback control real-time magnetometers have been reported,[16,18,24,25] diamond NV magnetometer with a high tracking rate remains a challenge. Based on the conventional discrete laboratory system, near-field magnetometer using a single NV center achieves a sensitivity of 6 μT·Hz$^{-1/2}$ and the calculated tracking rate is 10 mT/s.[16] Similarly, a real-time peak-locking method has been proposed for magnetic field detection at low photo count rates, whereby a sensitivity of 4.1 μT·Hz$^{-1/2}$ and tracking rate of 50 μT/s are reached.[24] The key point to the limited tracking rate is mainly due to the millisecond level switching time of the microwave source.

In this study, we demonstrate the direct measurement of a fast-changing magnetic field with an ensemble of NV centers in a diamond-based on board-level integrated system. Using a closed-loop frequency locked (CLFL) scheme, we construct a novel system to track the time-variant external magnetic field with an extended dynamic range far beyond the NV resonance. By reducing the microwave (MW) switching and demodulation time in the loop, the measured bandwidth is enhanced up to 10 kHz, enabling a maximum tracking rate of 0.723 T/s. We also investigate the sensitivity and performance of the CLFL system using single-frequency modulation (SFM) and three-frequency modulation (TFM). The difference is the use of single or three frequency microwaves for driving population transfer of electron spin states. A higher tracking rate is obtained with the SFM due to the trade-off between tracking performance and sensitivity. Our board-level system is not only compact and portable, in contrast to bulky lab-based equipment, but also demonstrates easy use under complex circumstances.

## II. FUNDAMENTAL AND SYSTEM DESIGN

The basic principle of NV-center magnetic-field measurement depends on the energy level structure of the NV center [Fig. 1(a)].[26,27] The electron spin states of the $S=1$ system are denoted by the spin quantum numbers $m_s=0, \pm 1$. Electrons excited by a 532 nm pump laser can decay in two paths. The electron spins in the $m_s=0$ excited state have a higher probability than those in the $m_s=\pm 1$ in generating red fluorescence (637-800 nm). A resonant MW field enables population transfer between the $m_s=0$ and $m_s=\pm 1$ states, indicating that the spin states can be determined by ODMR measurement. At room temperature, the zero-field spin-spin splitting ($D_{gs}$) from $m_s=0$ to $m_s=\pm 1$ is approximately 2.87 GHz. In an applied magnetic field, the spin levels are split due to the Zeeman effect and are governed by:[26]

$$f_{0\pm} \approx D_{gs} + \beta \Delta T \pm \gamma B_{NV} \tag{1}$$

where the temperature coefficient $\beta \approx -74$ kHz/K,[28] NV gyromagnetic ratio $\gamma = g_e \mu_B/h \approx 28$ Hz/nT, and $\Delta T$ is the temperature offset.[26] The energy level resonance frequency is further split by 2.16 MHz due to the hyperfine interaction between the electron spin and $^{14}$N nuclear spin ($I = 1$).

Figure 1(b) depicts the basic operation principle of the real-time tracking of an external magnetic field using the CLFL technique.[16] In the ODMR measurement, frequency modulation is applied to the MW source, and a dispersion type curve of the spin resonance can be measured using a lock-in amplifier (LIA). Different ODMR line-shapes are obtained by switching the modulation mode between the SFM and TFM mode.[29] A change in the external magnetic field causes a shift in the resonance frequency i.e. the center frequency (controlled by CLFL) now becomes the new resonance frequency. This ensures that the external magnetic field is invariably located within the dynamic range.[5,6,30]

Figure 1(c) demonstrates the CLFL measurement system. A conventional MW signal generator is not reliable to rapidly track the NV resonance frequency as its frequency switch time is around a few milliseconds. Thus, this work utilizes synthesized frequency-agile MW in which the frequency can be switched rapidly. A schematic of the frequency-agile MW generation is illustrated as the green block in Fig. 1(c). Direct digital synthesis (DDS) programmed in a Xilinx field-programmable gate array (FPGA) is used to generate a frequency modulated signal $f_{FM}$ with a carrier frequency $f_0$=9 MHz, modulation frequency $f_{mod}$=30 kHz, and a modulation depth of $f_{dev}$. A square-wave modulation is employed as it has the highest slope in the ODMR curve.[29] Frequency modulated MW signal $f_1$ is obtained by mixing a local oscillator (ADI, HMC833) signal $f_{LO}$ with frequency modulated signals $f_{FM}$ using an IQ mixer (Marki Microwave, IQ-1545). Then, a frequency-agile source (Novatech, DDS8P) generates a radio frequency signal ($f_{agile}$) with an ultra-short-frequency switching time (<600 ns). Frequency hopping is directly controlled by the FPGA through a 48-bit parallel port. Similarly, the MW signal for spin manipulation is synthesized by mixing $f_1$ and $f_{agile}$ in a single-sideband (SSB) mixer to suppress the image frequency signal by implementing a phase-cancellation technique. The final synthesized agile MW frequency $F_a$ is given by $F_a=f_{LO}+f_0+f_{dev}sgn[sin(2\pi f_{mod}t)]+f_{agile}$.

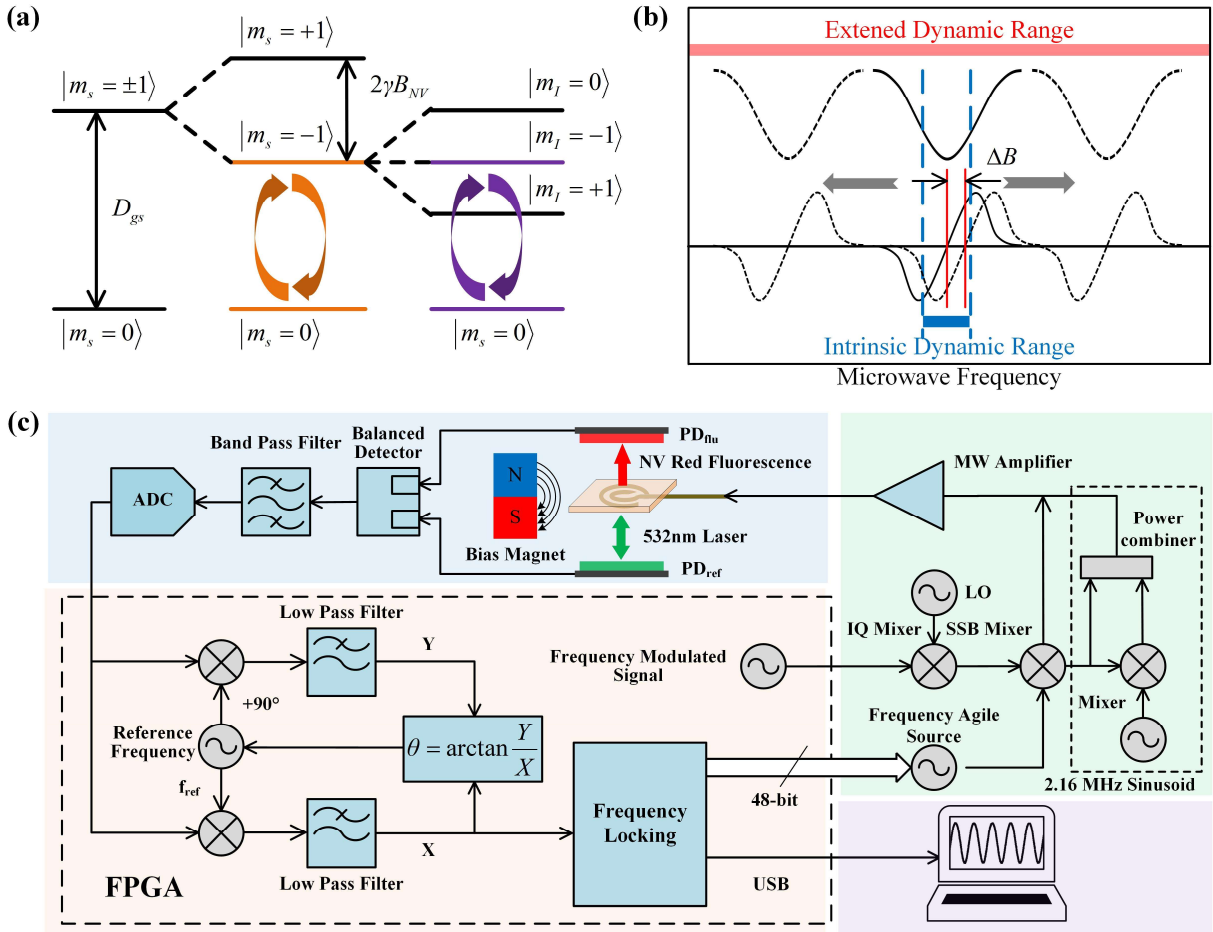

FIG. 1. (a) Ground-state energy-level diagram of the NV center. (b) Basic principle of closed-loop frequency-locked magnetic measurements. (c) Block diagram of the closed-loop frequency-locked system. The green block represents the microwave module, the blue block the magneto-opto-electrical conversion module, the yellow block the digital signal processing module, and the purple block the upper computer.

In SFM-ODMR mode, $F_a$ is amplified using an MW power amplifier (Mini-Circuits, ZHL-16W-43-S+). To achieve TFM-ODMR,[31] $F_a$ is mixed with a 2.16 MHz-sinusoid waveform using a double balanced mixer (Marki MW, MM1-0212L) to obtain two sidebands. Three-tone MWs are generated by combining $F_a$ with the two sidebands using a power combiner (Talent Microwave, RS2W0560-S).

A diamond is mounted on a coplanar MW waveguide used for MW delivery. In this work, the diamond is a single crystal of dimensions 3×3×0.5 mm$^3$, with a concentration of [N]<13 ppm, grown by chemical vapor deposition from Element Six Ltd. A 0.3 W 532 nm laser (Coherent Verdi 6) is used to excite the NV centers. The laser passes through a beam splitter, in which one part is focused on the surface of the diamond through a 10× objective lens (Nikon, Plan Apo λ, NA=0.45), and the other part of the laser is directed to the reference arm of a balanced detector (Thorlabs, PDB450A) to cancel out the laser-intensity noise. The NV centers are continuously excited by both the resonance MW field and the focused laser beam simultaneously. The red fluorescence emitted by the NV color centers then passes through a long-pass filter and is directed onto the signal arm of the balanced detector. The typical power of the collected fluorescence is ~23 μW, corresponding to a photocurrent of 10.6 μA. After band-pass filtering, a 12-bit analog-to-digital converter (ADI, AD9238) converts the analog signal into a digital signal, which is demodulated by the FPGA.

An LIA is programmed on the FPGA platform. First, the signals collected by the analog-to-digital converter (ADC) are sent to the digital processing module, where they are multiplied by the reference signal generated internally by the DDS. A low-pass filter is then implemented to reduce high frequency noise. The resulting signal is defined as in-phase demodulation $X$. To mitigate against the reduction in the signal-to-noise ratio, caused by the phase difference $\theta$ between the input and the reference signal, quadrature demodulation (QD) is carried out, which is similar to in-phase demodulation except that the reference signal has 90° phase shift. The QD output signal $Y$ and phase difference $\theta$ is given by $arctan(Y/X)$, for which the phase difference can be zeroed by compensating for the phase $\theta$ of the reference signal to achieve the optimal signal-to-noise ratio.

A non-zero signal $\Delta V$ demodulated by LIA is used to calculate the magnetic-field change $\Delta B$ and frequency shift $\Delta f$, given by $\Delta B=\Delta V/k\gamma$ and $\Delta f=\Delta V/k$, where $k$ is the slope at the zero-crossing position of the resonance curve. With a frequency-agile MW source of bandwidth $F_{band}$ =120 MHz, the dynamic range of the system is bound by $F_{band}/\gamma$, indicating a dynamic range of 4.3 mT. In each measurement cycle, a set of real-time magnetic-field changes and frequency shifts are calculated and displayed on a PC terminal. The real-time resonance frequency is controlled by a 48-bit parallel port handshake protocol to complete the whole closed-loop measurement cycle. A photograph of the home-built FPGA-based CLFL system is shown in Fig. 2(a).

In our experimental set up [Fig. 2(b)], a permanent magnet is placed ~5cm from the diamond, such that the bias field at the diamond is ~12 mT. This diamond is placed at the center of a custom-built double-layer Helmholtz coil, driven by an adjustable current to produce a fast-changing magnetic field.

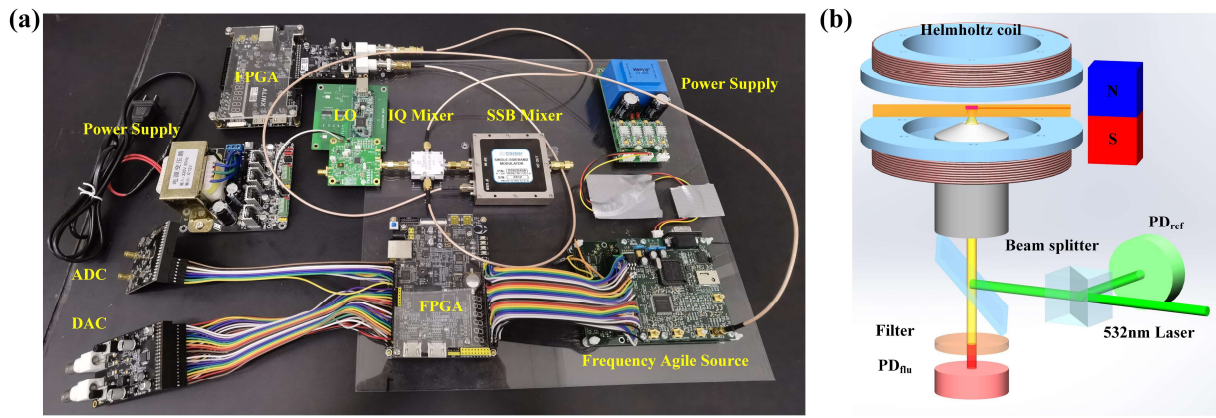

FIG. 2. (a) Home-built FPGA-based closed-loop frequency-locked system. (b) Schematic of the NV-center magnetic-measurement setup.

## III. RESULTS

### A. Properties of the synthesized microwave signals

The synthesized frequency agile MWs were characterized based on the frequency switching time and the sideband and carrier suppression performances. As seen in Fig. 3(a), it takes ~36 ns to switch the frequency from 11 MHz to 51 MHz. This is negligible compared to the demodulation time of the LIA (more than tens of μs). Empirically, the crosstalk of the ODMR signals of the undesired spin transitions can be eliminated with a suppression ratio of an undesired sideband of greater than 30 dBc. In Fig. 3(b), the FFT spectrum of the synthesized agile MW signal displays the sideband and carrier suppression at around 3.02 GHz is 30 dBc. Thus, most of the MW power is used to drive spin transitions.

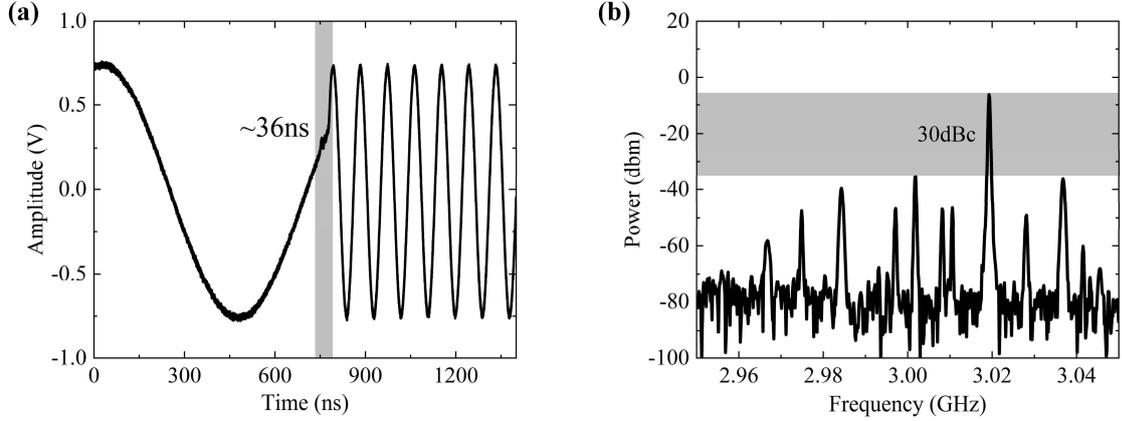

FIG. 3. Characterization of synthesized frequency agile microwave signal. (a) Frequency switching of frequency agile source controlled using a 48-bit parallel communication interface. It takes 36 ns for the frequency to switch from 11 MHz to 51 MHz. (b) Spectrum of the synthesized agile MW signal delivered to the antenna for spin manipulation. The sideband and carrier suppression are higher than 30 dBc.

## B. Magnetic-field-tracking performance

Magnetic-field tests to study the magnetic tracking performance of different ODMR modalities were carried out with SFM and TFM-OMDR with a laser and MW power of 0.3W and 19.8 dBm, respectively. In SFM, the modulation deviation $f_{dev}$ is set to 3 MHz to obtain an ODMR curve with an optimized slope and a relatively large linear region;[29] the intrinsic dynamic range is ~264 μT, and the maximum slope of linear fit at the resonance point is 0.28 μV/Hz [Fig. 4(a)]. The magnetic noise density spectrum indicates a mean noise floor of 10.5 nT·Hz$^{-1/2}$. A peak at 50 Hz is observed at the resonance point corresponding to the ambient magnetic noise generated by laboratory instruments [Fig. 4(c)]. Alternatively, for TFM, the modulation depth $f_{dev}$ is set to 500 kHz. Five resonance peaks are observed in the TFM-ODMR spectrum. The slope of the innermost zero crossing is the steepest due to the contribution from three MW frequencies resonantly addressing all three of the hyperfine features, simultaneously. An enhanced slope is obtained at 1.08 μV/Hz [Fig. 4(b)]. As a result of hyperfine splitting, we observe a higher sensitivity of 4.2 nT·Hz$^{-1/2}$ with the caveat that the intrinsic dynamic range is reduced to 50 μT [Fig. 4(d)]. To evaluate the overall performance of the FPGA-based board-level system, we characterize the sensitivity using discrete lab-based systems for comparison. For example, a MW source (Stanford, SG396), ADC module (NI, PCIE 6374), and LIA (Stanford, SR850) are employed for MW control and signal processing. The measured sensitivity of the discrete lab-based system is found to be an order of magnitude higher than the FPGA-based board-level system [Fig. 4(e) and (f)]. Optimized electric designs for preprocessing, such as band pass filtering and low noise amplification of LIA, could be used for noise suppression and sensitivity improvement.

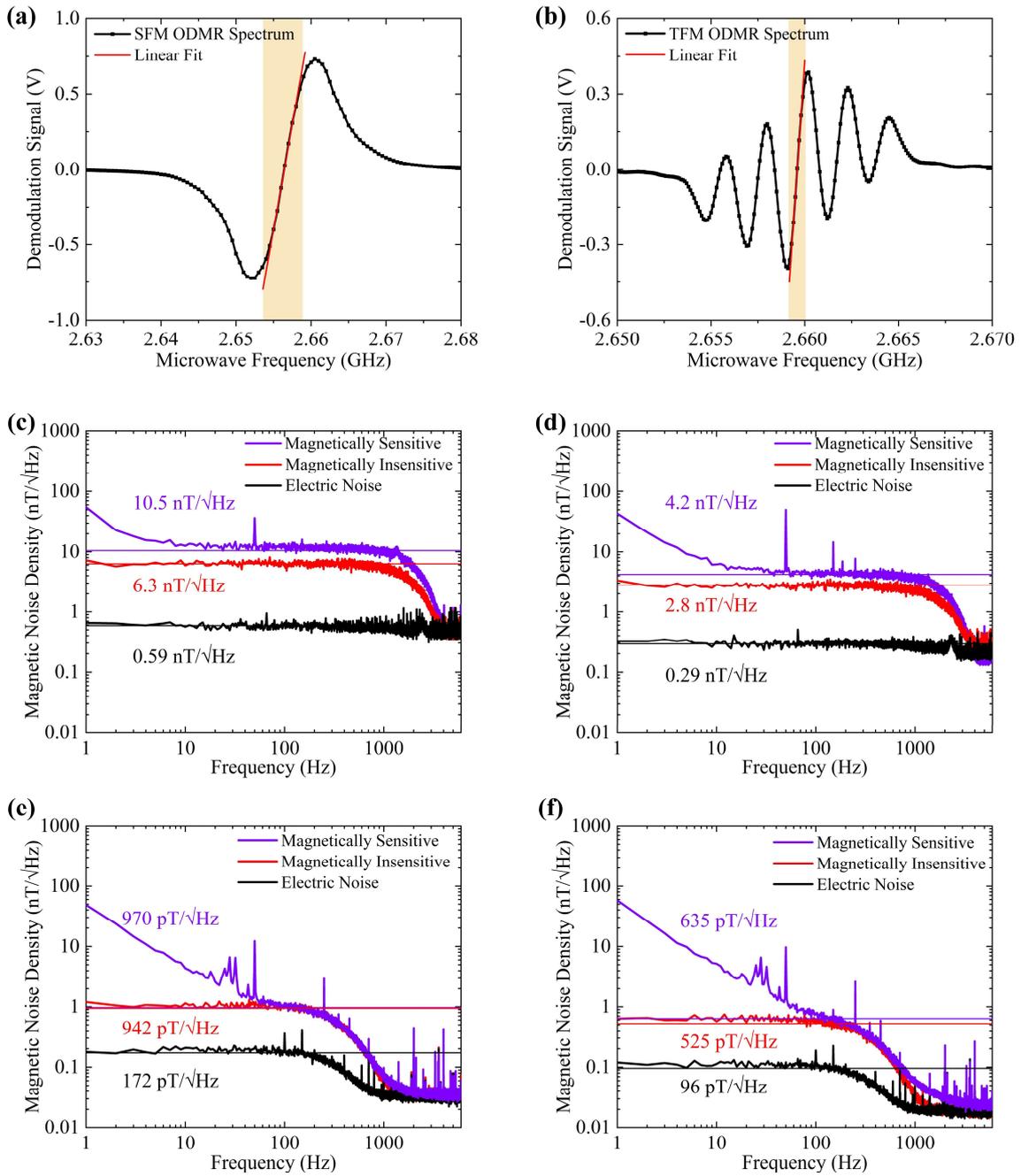

FIG. 4. (a) SFM and (b) TFM-ODMR spectra, respectively. The yellow shaded area represents the intrinsic dynamic

range. (c) SFM and (d) TFM magnetic noise density spectrum of the FPGA-based board-level system. Mean noise floors for SFM and TFM exist at 10.5 nT·Hz$^{-1/2}$ and 4.2 nT·Hz$^{-1/2}$ with laser noise rejection, respectively. (e) SFM and (f) TFM magnetic noise density spectrum obtained using discrete lab-based systems. Mean noise floors for SFM and TFM exist at 942 pT·Hz$^{-1/2}$ and 525 pT·Hz$^{-1/2}$ with laser noise rejection, respectively.

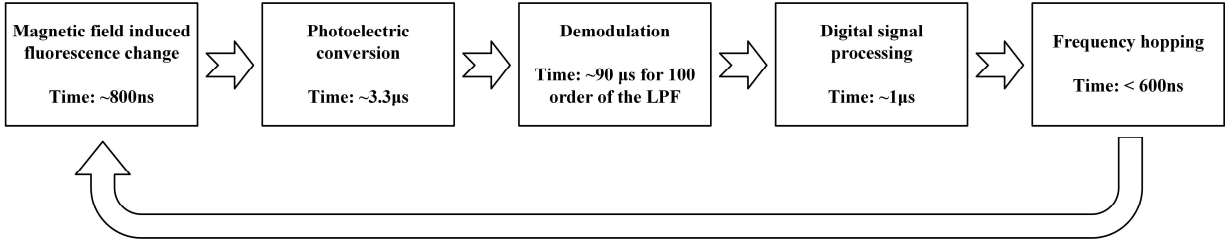

FIG. 5. Flowchart of each step of the closed-loop frequency-locked cycle with their respective processing times. In ODMR measurement mode, the photon counts of the spin state $m_s=0$ and $m_s=\pm1$ is stable after ~800 ns of laser pumping. The bandwidth of the photodiode detector is 0.3 MHz with a response time of 3.3 μs. An ADC with a 60 MHz sample rate converts analog signals to digital signals, and an FPGA-based lock-in amplifier demodulates the fluorescence signal. In each cycle, the time of demodulation mainly depends on the order of the digital low-pass filter; for example, the demodulation time of the 100-order filter is ~90 μs. Digital signal processing, including magnetic-field and resonance-frequency calculations and command encoding for frequency hopping, takes ~1 μs. According to the resonance frequency shift, the frequency hopping is controlled by a 48-bit parallel communication interface with a switching time of less than 600 ns.

Next, we study the relationship among the measurement bandwidth, the magnetic-field-tracking rate, and the measurement accuracy. The theoretical magnetic-field-tracking-rate limit $v_{Bmax}$ can be expressed as:

$$v_{Bmax} = \frac{\Gamma}{2t_{cycle}}, \qquad (2)$$

where $\Gamma$ is the intrinsic dynamic range, and $t_{cycle}$ is the time for one closed-loop cycle. Once the variation in the magnetic field in a closed-loop period exceeds half that of the intrinsic dynamic range, $\Gamma/2$, the CLFL system breaks out of the loop. The key and most time-consuming component of the LIA for demodulation is the finite impulse response low-pass filter since it is not only used to filter out the high-frequency components of the incoming data but also retain the direct current (DC) and some of the low-frequency components. As $t_{cycle}$ increases with the order of filter, conversely, the measurement bandwidth $1/t_{cycle}$ decreases with the order of filter. As shown in Figs. 6(a) and (b), both the maximum magnetic-field-tracking-rate and the measured standard deviation of magnetic field increase with measurement bandwidth. Figure 5 shows flowchart of CLFL cycle. The fluorescence takes ~800 ns to reach stability after the external magnetic field has changed.[32] For a detector with a photoelectric conversion time of 3.3 μs, the rise time of the signal processed by a 100-order low-pass filter is ~90 μs. To obtain a stable demodulation signal and avoid overshooting, the output data is extracted after 8 μs, hence leaving the sampling interval of the demodulation signal at 98 μs. Since the mathematical operations and the subsequent frequency switching time take less than 2 μs, a bandwidth of 10 kHz is achieved. Therefore, the theoretical magnetic-field-tracking-rate limits of SFM and TFM are 1.32 T/s and 0.25 T/s, respectively. A higher bandwidth is not pursued since it is difficult to stabilize the frequency-locked magnetic measurement resulting in an increase in burr noise. Figure 6(c) represents an extreme case for tracking in SFM mode with a 325 Hz sinusoidal magnetic field with an amplitude of 360 μT. Hence, the tracking rate is calculated to be 0.723 T/s with a standard deviation of 2.86 μT [Fig. 6(b)]. In comparison, the maximum tracking rate is 0.103 T/s with a standard deviation of magnetic field of 0.73 μT in TFM-ODMR mode [Fig. 6(d)]. When the measurement bandwidth is fixed, the intrinsic dynamic range limits the maximum field change in each cycle and further impeding the maximum magnetic-field-tracking rate. However, the measurement bandwidth can be enhanced to achieve a higher tracking rate, until the fluctuation induced by the noise exceeds the intrinsic dynamic range.

Next, the step response is analyzed with a measurement bandwidth of 10 kHz. Figure 6(e) shows the tracking of a 75 Hz square-wave-driven magnetic field with an amplitude of 390 μT, which is the extreme case for the SFM mode. The rising and falling magnetic tracking rates are 0.712 T/s and 0.641 T/s respectively. Similarly, the maximized tracking rates for rising and falling are 0.095 T/s and 0.090 T/s, respectively, under a 75 Hz square-wave-driven magnetic field with an amplitude of 46 μT for TFM [Fig. 6(f)]. A slow ramp-up of the magnetic field is observed due to the self-induction of the Helmholtz coil.

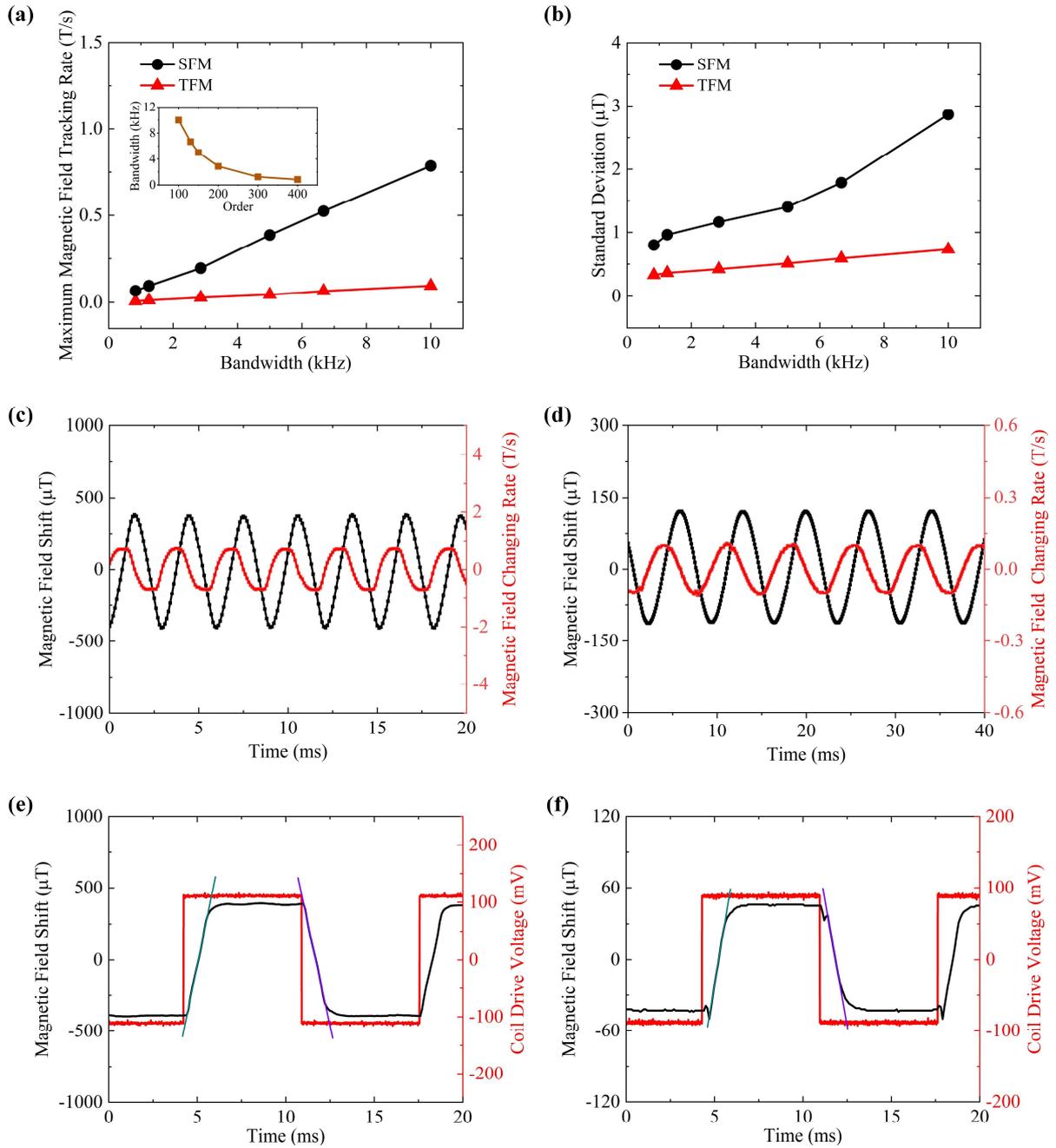

FIG. 6. (a) Plot of continuous magnetic-field-tracking rate and (b) measured standard deviation of magnetic field as a function of measurement bandwidth for SFM and TFM-ODMR. The inset in (a) is a plot of bandwidth as a function of the order of the low-pass filter. Plots of maximum attainable varying magnetic field and derived slope as functions of time for SFM (c) and TFM-ODMR (d). The corresponding tracking rates are 0.723 T/s and 0.103 T/s for SFM and TFM-ODMR, respectively. The square-wave signal generates abrupt rising and falling. (e) and (f) Time traces of the maximum attainable square-wave magnetic fields for SFM and TFM-ODMR, respectively. Linear fits are used to calculate the tracking rates, which are 0.712 T/s and 0.095 T/s for SFM and TFM-ODMR, respectively.

**C. Dynamic range**

Based on the working principle of CLFL, the dynamic range of magnetic-field measurement can be extended in accordance with the bandwidth of the frequency-agile source. To verify the extended dynamic range, a sinusoidal magnetic field is generated by the Helmholtz coil with a peak-to-peak value of the magnetic field reaching 3.8 mT in projection of the NV axis of interest. Again, the initial bias magnetic field is determined and agile MWs are located at the resonance frequency. The varying magnetic field is measured in a 10 kHz bandwidth by the CLFL system with an amplitude of ~3.8 mT and a frequency of 41 Hz [Fig. 7], corresponding to a resonance frequency shift of ~106.4 MHz.

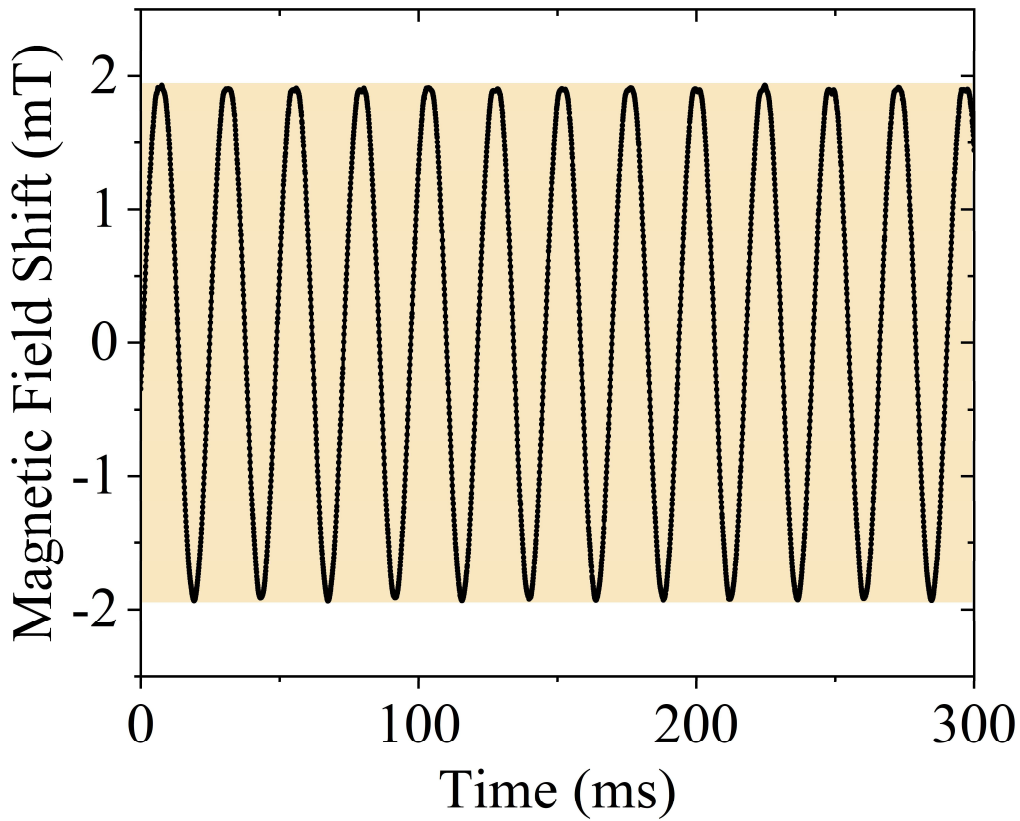

FIG. 7. Time trace of 41 Hz alternating current (AC) magnetic-field shift with extended dynamic range of ~3.8 mT produced by a Helmholtz coil.

### D. Linearity

Since linearity is one of the key parameters in magnetometers, we investigated the nonlinear errors for both the AC and DC modes of the diamond NV magnetometer. Figure 8 renders the linearity measurement results for the SFM and TFM modes at a measurement bandwidth of 10 kHz. The time traces of the AC magnetic fields are shown in Figs. 8(a) for SFM mode and 8(b) for TFM mode. The time traces of the DC magnetic fields are shown in Figs. 8(c) for SFM mode and 8(d) for TFM mode. Figure 8(e) shows plots of the AC and DC magnetic-field amplitudes as functions of the driving current with derived non-linearity errors of 0.217% and 0.463% in SFM mode. Figure 8(f) portrays the plots of the AC and DC magnetic-field amplitudes as functions of the driving current with derived non-linearity errors of 0.268% and 0.306% in TFM mode. In measuring the AC magnetic field, the driving current value captured by an ammeter is the root-mean-square (rms) value of the AC current. Therefore, this causes a discrepancy between the slope values of the AC and DC magnetic fields in Figs. 8(e) and 8(f). The AC magnetic-field measurement demonstrates a better linearity than the DC measurement. This is because the DC magnetic-field measurement is more affected by temperature drift.[33] Moreover, the two-channel CLFL system can be used to mitigate the temperature drift, thus improving the linearity.[34] The frequencies of the AC magnetic fields are 10 Hz for SFM and 30 Hz for TFM.

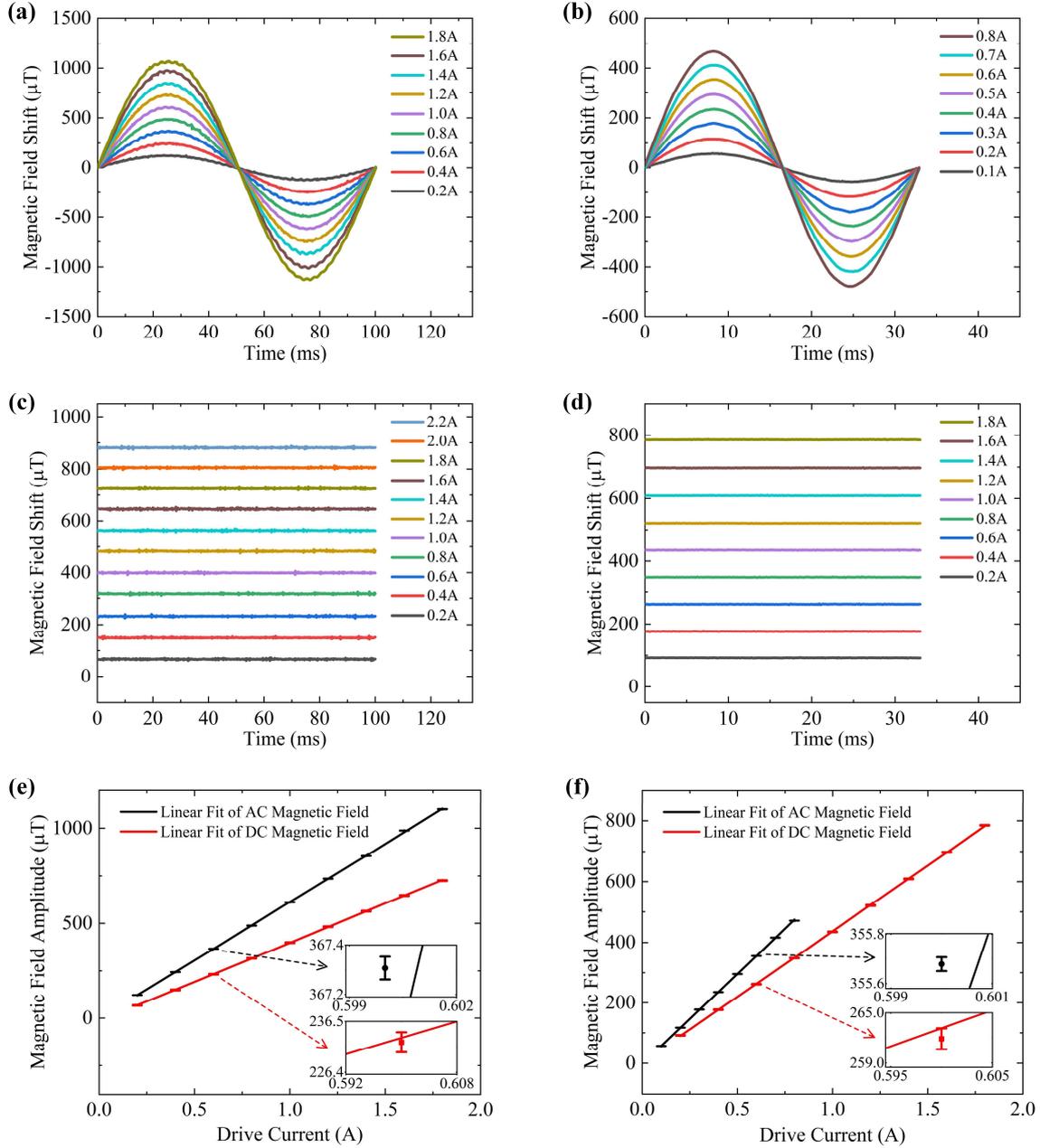

FIG. 8. Linearity characterization of AC and DC magnetic-field measurements. Measured magnetic fields generated by AC currents (rms) projected on the NV axis using (a) SFM and (b) TFM-ODMR. Measured magnetic fields with discrete DC currents using (c) SFM and (d) TFM-ODMR. Plots of amplitude for (e) AC and (f) DC magnetic fields as functions of driving current. The linearity of AC and DC magnetic-field measurement reaches 0.217% and 0.463% for SFM-ODMR, respectively. In contrast, the linearity of AC and DC magnetic-field measurement reaches 0.268% and 0.306% for TFM-ODMR, respectively. We obtain AC magnetic field values of 10 Hz and 30Hz for SFM and TFM,

respectively, when a 10 kHz measurement bandwidth is applied for linearity characterization.

E.  Application

To confirm the applicability of the NV magnetometer, we measured the power supply of a soldering iron. The power cable of the electric soldering iron is situated ~20 cm away from the NV magnetometer. For high-accuracy measurements, TFM-ODMR mode with a measurement bandwidth of 1.25 kHz is imposed. When the electric soldering iron is turned on, an oscillating field with a dominant frequency of 50 Hz and its harmonic are observed [Fig. 9]. To combat the numerous burrs produced in the absence of an external magnetic field, the shielding of the CLFL system electronic connector can be improved to optimize future electrics design.

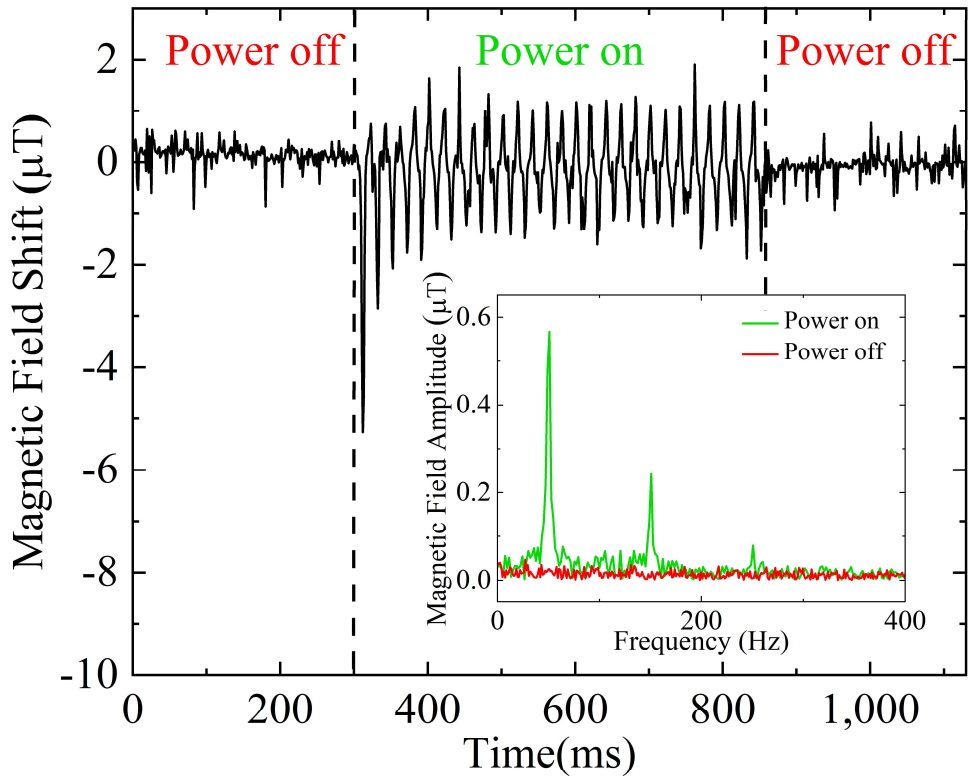

FIG. 9. Time traces of measured magnetic field induced by switching an electric soldering iron on. The power cable of the electric soldering iron is placed 20 cm away from the NV magnetometer. The measurement bandwidth of the closed-loop frequency-locked system is set to 1.25 kHz. The NV magnetometer monitors the power supply of the electric soldering iron in real time. Once the iron is turned on, a 50 Hz AC magnetic field and its higher-order harmonic are observed in the frequency domain (inset).

## IV. CONCLUSION

We present a fast-changing magnetic-field-tracking method based on an ensemble of nitrogen vacancy centers in diamonds, along with board-level integrated electronics. To combat the limitations of intrinsic dynamic range in open loop measurement, a closed-loop frequency-locked method is implemented in which the measurement bandwidth is greatly increased up to 10 kHz. The magnetic tracking performances for both SFM and TFM optically detected magnetic resonance modes are analyzed with a maximum tracking rate of 0.723 T/s achieved. Moreover, temperature drift correction and scale-factor-free vector magnetic-field measurement can be realized by simply increasing the number of closed-loop frequency-locked channels. Integration of both the compact diamond sensor devices[20] and demodulation system pave the way toward practical applications by transforming lab-based instruments into highly integrated compact devices. Additionally, the tracking rate and dynamic range of closed-loop frequency-locked system can be further improved for other applications such as low-field MRI scanners and precise, high-current sensors.

**ACKNOWLEDGMENTS**

The authors acknowledge support from the CAS Strategic Pilot Project (No. XDC07030200), R&D Program of Scientific Instruments and Equipment, Chinese Academy of Sciences (No. YJKYYQ20190026), and the National Key R&D Program of China (No. 2021YFB3202500).

**DATA AVAILABILITY**

The data that support the findings of this study are available on request from the corresponding author. The data are not publicly available due to state restrictions.